\documentclass {jetpl}
\twocolumn

\lat

\title {Electronic structure and magnetic state 
of transuranium metals \\ under pressure}

\rtitle {Electronic structure and magnetic state of transuranium metals under pressure}

\sodtitle {Electronic structure and magnetic state 
of transuranium metals under pressure}

\author {A.\,V.\,Lukoyanov$^{+,*}$, A.\,O.\,Shorikov$^{+}$, V.\,B.\,Bystrushkin$^{*}$, 
A.\,A.\,Dyachenko$^{*}$, L.\,R.\,Kabirova$^{*}$, Yu.\,Yu.\,Tsiovkin$^{*}$, A.\,A.\,Povzner$^{*}$, 
V.\,V.\,Dremov$^{**}$, M.\,A.\,Korotin$^+$, V.\,I.\,Anisimov$^{+}$ 
}

\rauthor {A.\,V.\,Lukoyanov, A.\,O.\,Shorikov, V.\,B.\,Bystrushkin et al.}

\sodauthor {A.\,V.\,Lukoyanov, A.\,O.\,Shorikov, V.\,B.\,Bystrushkin, 
A.\,A.\,Dyachenko, L.\,A.\,Kabirova, Yu.\,Yu.\,Tsiovkin, A.\,A.\,Povzner, 
M.\,A.\,Korotin, V.\,V.\,Dremov, V.\,I.\,Anisimov}

\address {$^+$Institute of Metal Physics, Russian Academy of Sciences--Ural Division, 
620990 Yekaterinburg, Russia\\~\\
$^*$Ural Federal University--UPI, 620002 Yekaterinburg, Russia\\~\\
$^{**}$Russian Federal Nuclear Center, Institute of Technical Physics, 
Snezhinsk, 456770 Chelyabinsk Region, Russia}

\dates {\today}{*}

\abstract {Electronic structure of bcc Np, fcc Pu, Am, and Cm pure metals 
under pressure has been investigated employing the LDA+$U$ method with 
spin-orbit coupling (LDA+$U$+SO). Magnetic state of the actinide ions 
was analyzed  in both $LS$ and $jj$ coupling schemes to reveal the applicability 
of corresponding coupling bases. It was demonstrated that whereas Pu and Am 
are well described within the $jj$ coupling scheme, Np and Cm can be described 
appropriately neither in $\{m\sigma\}$, nor in $\{jm_j\}$ basis, due to 
intermediate coupling scheme realizing in these metals that requires some 
finer treatment. The LDA+$U$+SO results for the considered transuranium 
metals reveal bands broadening and gradual 5$f$ electron delocalization under pressure.}


\begin {document}
\maketitle

Prominent structural transition from $\delta$- to $\alpha$-phase of plutonium 
gives substantial total volume contraction~\cite{Lander03}. In americium 
and curium the similar transitions were also found 
under high pressure or increasing temperature~\cite{Lindbaum01,Heathman05}. 
These volume collapses are usually related to the drastic electron delocalization. 
In curium, additional magnetic stabilization of intermediate phases 
was also found~\cite{Heathman05,Moore07}.

In many strongly correlated materials Coulomb and Hund interactions are dominant, 
whereas spin-orbit (SO) coupling is comparably smaller and usually can be treated within 
various perturbative techniques. In the case of 5$f$ electronic shell of actinide 
metals all these three terms in Hamiltonian are of comparable strength. That results in 
a very sensitive (non)magnetic ground state, but also in coupling schemes 
varying from usual $LS$ (Russel-Saunders) one. An intermediate or $jj$ coupling 
schemes were found to be more appropriate for 5$f$ electrons 
in transuranium metals~\cite{Moore09,Moore03,Laan04,Laan96}.

In actinide elements spin-orbit coupling is stronger than exchange 
Hund interaction, and hence the $jj$ coupling scheme can be valid with 
a well defined total moment $\mathbf J$, but in this case spin $\mathbf S$ 
and orbital $\mathbf L$ moments are not well defined. Then the basis 
of eigenfunctions of total moment operator $\{jm_j\}$ is the best choice, 
since the matrix of spin-orbit coupling operator and occupation matrix 
are diagonal in this basis. However the exchange interaction (spin-polarization) 
term in the Hamiltonian is not diagonal.

In some 5$f$ elements intermediate coupling scheme is realized, then 
the occupation matrix has nondiagonal elements in both $\{m\sigma\}$ and $\{jm_j\}$ 
orbital bases. Therefore, both terms in the Hamiltonian: spin-orbit coupling and exchange 
interaction, should be taken in a general nondiagonal matrix form. 

Numerous band methods and approximations have been applied to describe magnetic 
and spectral properties of transuranium metals, see \cite{Moore09} for review. 
Nonmagnetic ground state of pure plutonium metal observed 
experimentally~\cite{Lashley05} was reproduced 
in the electronic structure calculations in the LDA+$U$+SO 
calculations~\cite{Shorikov05} (local density approximation supplemented 
with the Hubbard $U$-correction and spin-orbit coupling). In these calculations 
the exchange interaction was found to be the reason of artificial antiferromagnetic 
ordering in various LSDA+$U$ investigations. Also non-magnetic ground state was found 
in around-mean-field version of the LDA+$U$ method~\cite{Shick05}, and later on within 
LDA + Hubbard I approximation~\cite{Shick09}, and also hybrid density functionals 
with a dominant contribution of HF functional~\cite{Atta09}. 
Recently, the reliability of these results was supported by more detailed analysis 
of exchange interaction~\cite{Cricchio08}, and also in the LDA+DMFT 
calculations~\cite{Kotliar06}, supplementing LDA with the Dynamical Mean-Field 
Theory \cite{Georges96} (DMFT)~\cite{Savrasov01,Shim07,Zhu07,Marianetti08}.

While consistent interpretation of spectroscopic data is not found 
yet~\cite{Tobin08}, the LDA+$U$+SO method provides realistic magnetic 
state and allows one to estimate electrical resistivity in actinide 
metals~\cite{Tsiovkin07} and alloys~\cite{Tsiovkin09} in good agreement 
with experimental data. 

In this paper electronic structure of bcc Np, fcc Pu, Am, and Cm pure metals 
was calculated within the LDA+$U$+SO method introduced in detail in Ref.~\cite{Shorikov05}. 
In the method the exchange interaction 
(spin polarization) term in the Hamiltonian is implemented in a general nondiagonal matrix form 
regarding the spin variables. This form is necessary for correct description of 5$f$ electrons 
behavior for the cases of $jj$ and intermediate coupling types. 

\begin{figure}
\vspace{5mm}
\includegraphics[width=0.45\textwidth]{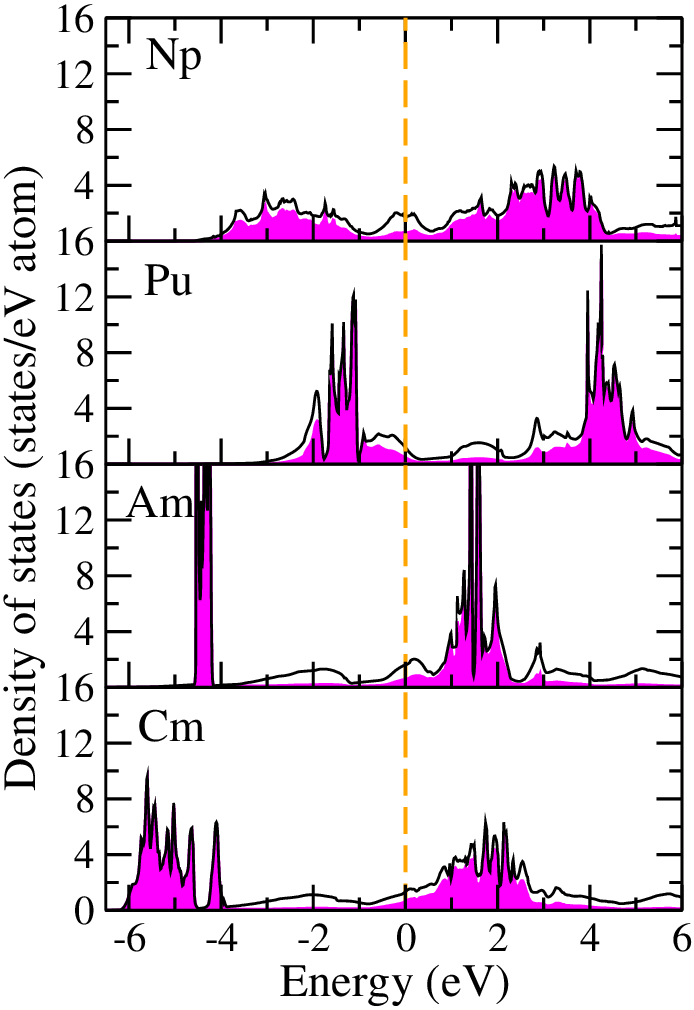}
\caption{FIG. 1. The density of states: total (solid line) and 5$f$ (shaded area) 
states of bcc Np, fcc Pu, Am, and Cm from the~LDA+$U$+SO calculation. 
The Fermi energy corresponds to zero.}
\label{fig:1}
\end{figure}

In the LDA+$U$+SO method accounting for strong electron correlations includes 
Coulomb interaction matrix elements which can be expressed via direct $U$ 
and exchange $J_H$ Coulomb parameters. A reliable way to estimate these parameters 
is provided by the constrain LDA approach~\cite{Gunnarsson89}. 
To calculate a value of $U$ parameter, in this procedure one evaluates 
a {\it screened} Coulomb interaction of 5$f$ electrons which 
requires the choice of screening channels taken into account. 
For $s$, $p$, and $d$ channels for bcc Np and fcc Pu, Am, and Cm 
the constrain LDA calculations resulted in the Coulomb parameter 
value $U$ = 4~eV ~\cite{Shorikov05,Savrasov01}.
The exchange Coulomb parameter $J_H$ is evaluated as the 
{\it{difference}} of interaction energy for the electrons pairs
with the opposite and the same spin directions. 
Parameter $J_H$ does not depend on the screening
channels choice. For neptunium and plutonium, the value 
of Hund exchange parameter $J_H$ was calculated as 
$J_H$~=~0.48~eV~\cite{Shorikov05}, for americium as 0.49~eV, 
and for curium as 0.52 eV~\cite{Tsiovkin07}. 
In this work in all LDA+$U$+SO calculations we used these exact values 
of $J_H$ for each actinide metal, since (non)magnetic state of actinide 
metals is sensitive to the value of $J_H$~\cite{Shorikov05}.

Strong spin-orbit coupling of 5$f$ electrons results in a splitting 
of the $f$ band into subbands corresponding to the values of total moment $j$=5/2 
and $j$=7/2. The value of this splitting is 1 -- 1.5 eV. 
Taking into account Coulomb correlations via the LDA+$U$ 
correction does not change qualitatively the band structure, 
only the separation between subbands increases from 1.5~eV
to 5~eV according to the value of $U$ = 4~eV. 

The density of states (DOS) from the LDA+$U$+SO calculations for Np, Pu, Am, and Cm 
at ambient pressure are shown in Fig.~\ref{fig:1}. In all DOSs one can distinguish two groups 
of bands: $j$=5/2 at the lower energies and $j$=7/2 at higher ones. In Np the $j$=5/2 
subband is partially filled. From Pu to Cm the Fermi level is shifted upward from 
the upper slope of $j$=5/2 subband in Pu and crosses the $j$=7/2 subband in Cm 
due to the increasing number of $f$ electrons. These results are for the metals 
at ambient pressure in the cubic phases. We used the same lattice parameters 
as in our previous work~\cite{Tsiovkin09}.

\begin{figure}
\vspace{5mm}
\includegraphics[width=0.45\textwidth]{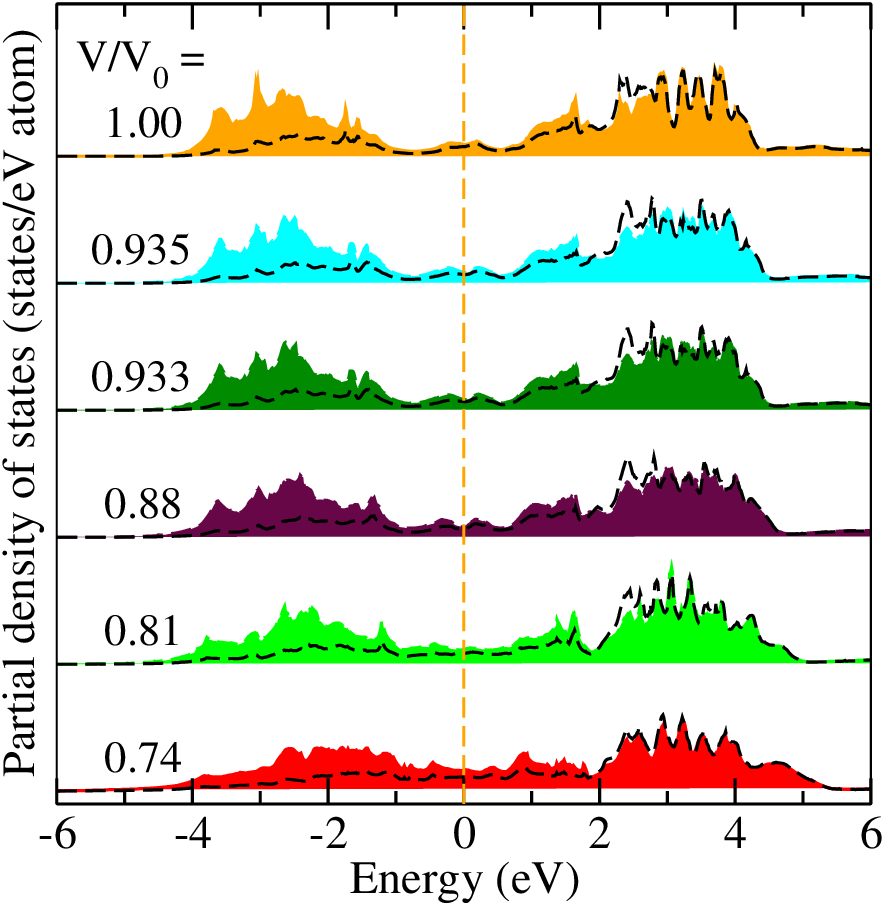}
\caption{FIG. 2. 
The partial DOS for the $j$=5/2 (shaded area) and $j$=7/2 (dashed line) 
subbands of bcc neptunium under pressure obtained from the LDA+$U$+SO calculations. 
For each curve the relative volume V/V$_0$ is given. The Fermi energy corresponds to zero.}
\label{fig:2}
\end{figure}

\begin{figure}
\vspace{5mm}
\includegraphics[width=0.45\textwidth]{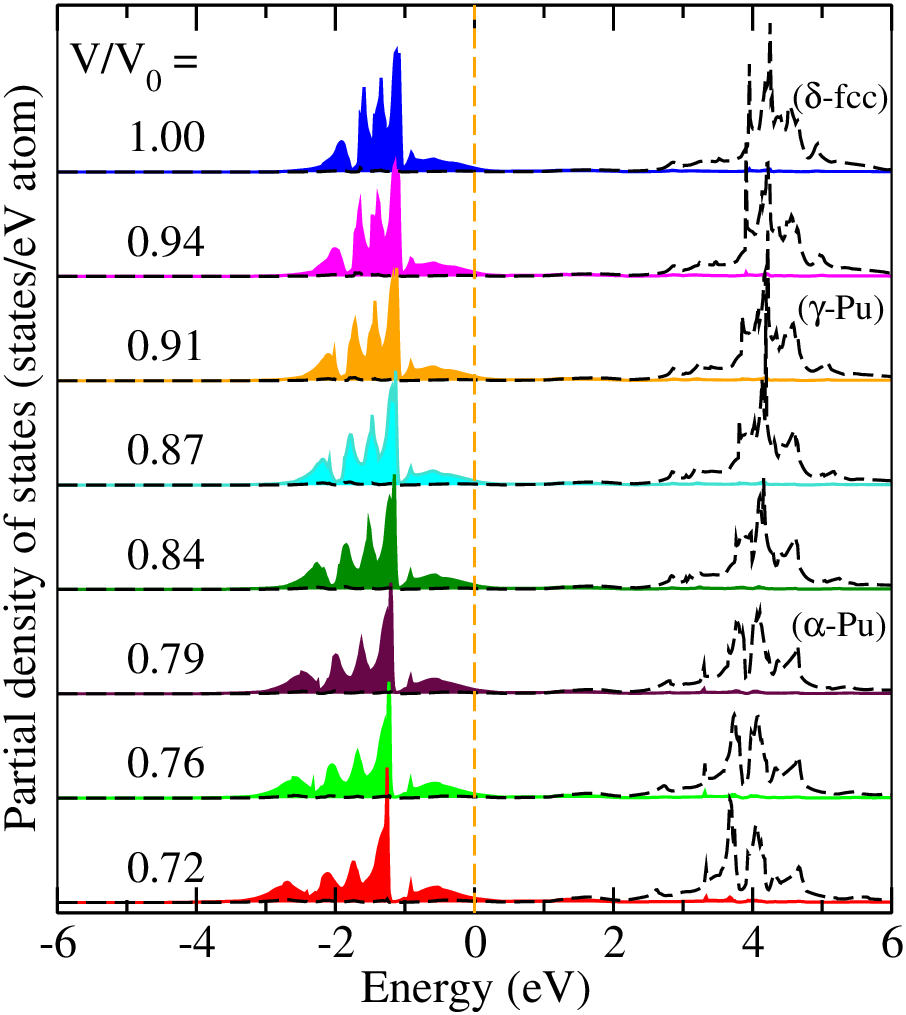}
\caption{FIG. 3. 
The partial DOS for the $j = 5/2$ (shaded area) and $j = 7/2$ (dashed line) 
subbands of fcc plutonium under pressure obtained from the LDA+$U$+SO calculations. 
For each curve the relative volume V/V$_0$ is given. Real phases with corresponding volumes 
per ion are cited in brackets. The Fermi energy corresponds to zero.}
\label{fig:3}
\end{figure}

\begin{figure}
\vspace{5mm}
\includegraphics[width=0.45\textwidth]{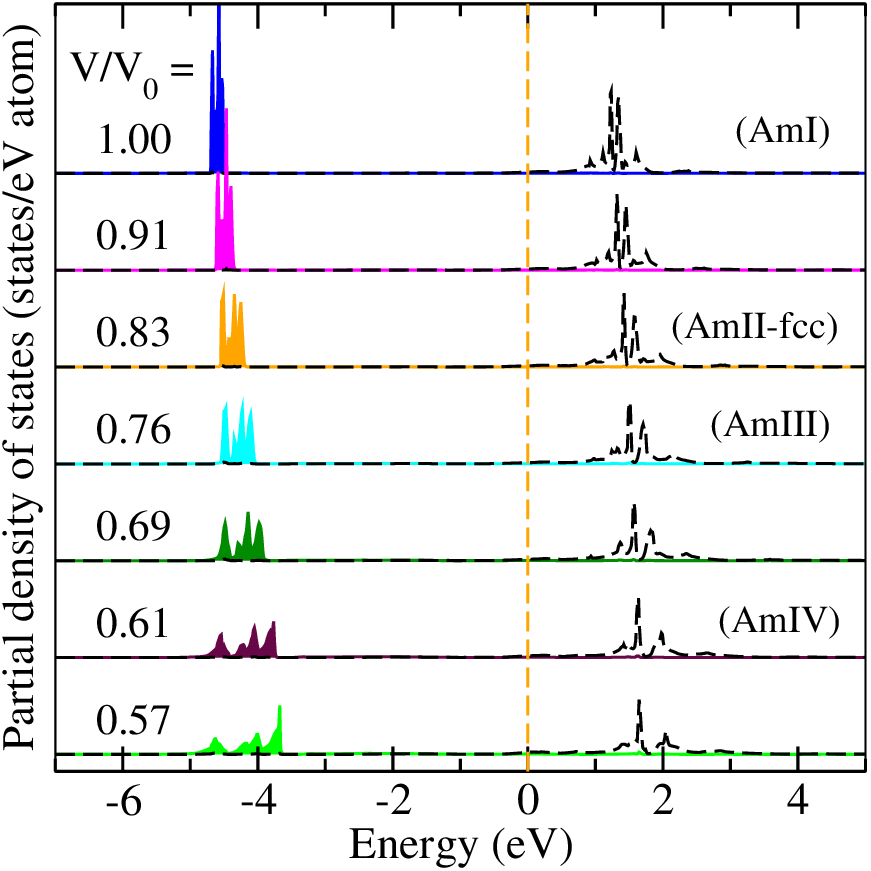}
\caption{FIG. 4. 
The partial DOS for the $j = 5/2$ (shaded area) and $j = 7/2$ (dashed line) 
subbands of fcc americium obtained from the LDA+$U$+SO calculations. 
For each curve the relative volume V/V$_0$ is given. Real phases with corresponding volumes 
per ion are cited in brackets. The Fermi energy corresponds to zero.}
\label{fig:4}
\end{figure}

\begin{figure}
\vspace{5mm}
\includegraphics[width=0.45\textwidth]{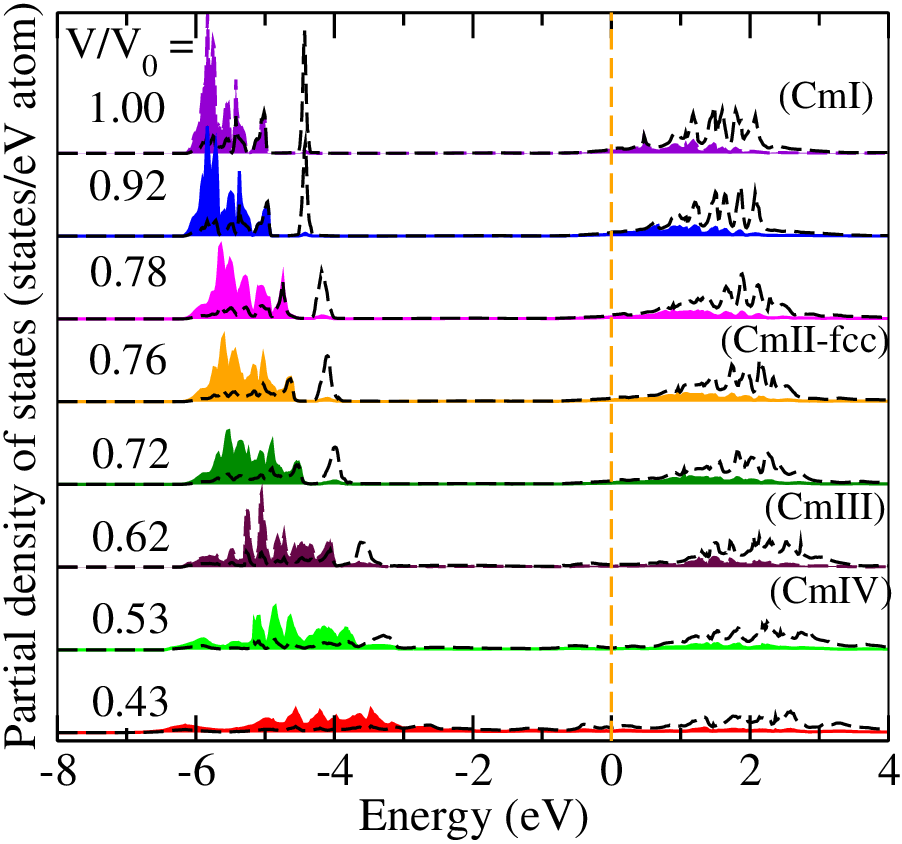}
\caption{FIG. 5. 
The partial DOS for the $j = 5/2$ (shaded area) and $j = 7/2$ (dashed line) 
subbands of fcc curium obtained from the LDA+$U$+SO calculations. 
For each curve the relative volume V/V$_0$, where V$_0$ is the volume 
at ambient pressure, is given. Real phases with corresponding volumes 
per ion are cited in brackets. The Fermi energy corresponds to zero.}
\label{fig:5}
\end{figure}

To model the pressure, we assume it to be uniform and applied as a cell volume contraction. 
While experimentally crystal structure of these metals transforms from cubic to complicated 
structures like orthorhombic and monoclinic, for the electronic structure and magnetic state 
of the actinide ion the main result of the applied pressure comes from the contraction of the unit cell 
volume per ion. In Figs.~\ref{fig:2}--\ref{fig:5} one can find density of states (DOS) for the cubic 
phases under pressure, real phases with corresponding volumes per ion are referred to in brackets.
In the actinide metals under investigation the cubic phases do not correspond 
to the largest volume per ion, for this reason we also consider some volumes exceeding 
the cubic volume at ambient pressure.

{\it Neptunium metal.}~--~
Magnetic ground state for Np at ambient pressure was found in our calculation (see Table 1). 
Large values of the off-diagonal elements (OD) in Table 1 in both $\{m\sigma\}$ 
and $\{jm_j\}$ bases evidence for intermediate coupling type in this 
actinide metal. Four largest eigenvalues close to the unit 
give $f^4$ configuration. 
Upon applied pressure the density of states at the Fermi energy increases, 
see Fig.~\ref{fig:2}, effective magnetic moment $\mu^{calc}_{eff}$ ranges from 2.55 $\mu_B$ 
to 2.33 $\mu_B$. 

{\it Plutonium metal.}~--~
The LDA+$U$+SO calculations for metallic Pu fcc structure (fcc phase in Pu is named 
$\delta$-phase) at all volumes gave a nonmagnetic ground state with zero values 
of spin $\mathbf S$, orbital $\mathbf L$, and total $\mathbf J$ moments~\cite{Shorikov05} 
in agreement with numerous experimental data~\cite{Lashley05}. 
The occupation matrix has six eigenvalues close to unit, see Table~1 (for 
details, see Ref.~\cite{Shorikov05}), 
and is nearly diagonal in the $\{jm_j\}$ basis of eigenfunctions 
of total moment $\mathbf J$. That gives a $f^6$ configuration of
Pu ion in $jj$ coupling scheme. In Fig.~\ref{fig:3} 
the partial densities of states for $f^{5/2}$ and $f^{7/2}$
subshell of Pu in all volumes (applied pressures) are presented. 
The DOSes  as well as  the occupation matrix demonstrate almost 
completely filled $f^{5/2}$ band with the Fermi level on the top of it 
and an empty $f^{7/2}$ band. The separation between the centers of these bands 
is $\approx$ 5.2 -- 5.4~eV. For small volumes $j$=5/2 subband does not change 
its position but its bandwidth becomes approx. 1 eV larger.

{\it Americium metal.}~--~
The calculates DOSes of Am are shown in Fig.~\ref{fig:4}. 
One can see that $j$=5/2 subband is fully occupied that corresponds 
to $f^6$ configuration whereas $j$=7/2 subband is almost empty. 
The occupation matrix has six largest eigenvalues of close to unit, see Table 1. 
The Fermi level in Am is shifted towards $j$=7/2 subband 
comparing to Pu due to additional valence electron (see Fig.~\ref{fig:4}) 
that occupies $s$, $p$, and $d$ states (not shown separately). 
Note, that the 5$f$ bandwidth in Am is smaller than in Pu due 
to the larger cell volume per ion. 
Having a delicate balance between SO and exchange interactions, calculated 5$f$-DOS 
occupied 5$f$ band is centered around 4 eV in agreement with Am photoemission 
spectra that demonstrate large density of states in the range (--4 eV; --2 eV)~\cite{Naegele84}. 
The LDA+$U$+SO calculations for Am  gave a nonmagnetic ground state in all volumes 
with the 5$f$ shell  with $S = L = J = 0$ in agreement 
with experimental data~\cite{Lindbaum01}.

\begin{table*}[!t]
\center{
\caption{TABLE 1. 
Electronic configuration of 5$f$ shell in Np, Pu, Am, and Cm ions in the cubic phases 
calculated within the LDA+$U$+SO method. The largest values of occupation matrices 
off-diagonal elements OD$_{LS}$ and OD$_{jmj}$ in the corresponding basis sets 
are given in the second and third columns. Then the seven largest eigenvalues 
of occupation matrix are presented. The columns from the eleventh to thirteenth 
show the calculated values for spin ($S$), orbital ($L$), total ($J$) moments.}}
\begin{tabular}{lllllll}
\hline 
Metal & OD$_{LS}$ & OD$_{jmj}$ & Largest eigenvalues & $S$ & $L$ & $J$\\
\hline
Np & 0.36 & 0.46 & 0.05 0.09 0.26 0.89 0.91 0.91 0.92 & 1.40 & 4.68 & 3.28\\
Pu & 0.45 & 0.01 & 0.03 0.92 0.92 0.93 0.93 0.93 0.93 & 0 & 0 & 0\\
Am & 0.47 & 0.02 & 0.07 0.97 0.98 0.98 0.98 0.99 0.99 & 0 & 0 & 0\\
Cm & 0.31 & 0.45 & 0.10 0.99 1.00 1.00 1.00 1.00 1.00 & 2.77 & 0.75 & 3.52\\
\hline 
\end{tabular}
\label{tab:1}
\end{table*}

{\it Curium metal.}~--~
In curium $f^{7/2}$ subband is partially filled, providing a peculiar empty state 
well below the Fermi level, see Fig.~\ref{fig:5}. The LDA+$U$+SO calculation results 
in $S$ and $L$ moments listed in Table 1. The value of effective magnetic moment 
$\mu_{eff}^{calc}$ calculated from the total moment value $J$ by the method 
used in Ref.~\cite{Shorikov05} 
for different volumes gradually decreases from 7.44 $\mu_B$ for the largest volume 
(corresponding to the volume per ion in the CmI phase) to 6.86 $\mu_B$ (CmIV). 
This result corresponds to the model calculations assuming in pure $LS$ 
coupling the magnetic moment of the curium ion to be 7.94 $\mu_B$, but for 
a realistic model of intermediate coupling lowering up to 7.6 $\mu_B$~\cite{Huray85}, 
while experimentally the magnetic moment was reported as 7.85 $\mu_B$~\cite{Huray85}. 

{\it In conclusion}, we have calculated the electronic structure 
of bcc Np, fcc Pu, Am, and Cm pure metals within the LDA+$U$+SO method 
under applied pressure in assumption of uniform unit cell contraction. 
Static mean-field approach used in this work results in gradual (featureless) 
electron delocalization and bands broadening under pressure. While $jj$ coupling 
scheme is well suited for description of Pu and Am ground state, 
for Np metal only intermediate coupling scheme seems to be appropriate, 
whereas in Cm the intermediate coupling is closer to LS-type. We suggest 
accounting for dynamical correlations effects will allow one to reproduce 
not only the delocalization of 5$f$ electrons under applied pressure 
but also will resolve some fine feature of this process.

This work was supported by the Russian Foundation for Basic Research 
(Projects Nos. 10-02-00046, 09-02-00431, and 10-02-00546), 
Federal Program  NK~529P, Russian Federal Agency for Science and Innovations 
(Program ``Scientific and Scientific-Pedagogical Training of the Innovating 
Russia'' for 2009-2010 years), grant No. 02.740.11.0217, the scientific program 
``Development of scientific potential of universities'' No.~2.1.1/779.

\end{document}